\documentclass[aps,twocolumn,showpacs]{revtex4}
\usepackage{graphicx}
\usepackage{epsfig}

\begin{document}
\title{Thermal noise for quantum state inference}

\author{D. Mogilevtsev$^{1,2}$, V. S. Shchesnovich$^{2}$, N. Korolkova$^{3}$}

\affiliation{$^1$Centro de Ci\^encias Naturais e Humanas,
Universidade Federal do ABC, Santo Andr\'e,  SP, 09210-170 Brazil;
\\
$^2$Institute of Physics, Belarus National Academy of Sciences,
F.Skarina Ave. 68, Minsk 220072 Belarus;
\\
$^3$School of Physics and Astronomy, University of St Andrews,
North Haugh, St Andrews KY16 9SS, UK;}

\begin{abstract}
In this work we describe a simple and efficient scheme for
inference of photon number distribution by adding variable thermal
noise to the signal. The inference remains feasible even if the
scheme parameters are subject to random dynamical change.
\end{abstract}


\maketitle

Quantum state tomography is the most advanced and complete
diagnostic tool available now. It allows one inferring maximal
possible information about the state of physical system or about
the process \cite{all}. Quantum tomography has already become a
standard experimental tool enabling reconstruction even such
fragile, exquisitely quantum objects as "Schrodinger cats"
\cite{Ourjoumtsev}. Intuitively, one expects that for performing
measurements able to collect information sufficient to reconstruct
a quantum state or process, it is necessary to build rather
precise measurement set-up. This intuition seems to be confirmed
by existence of limits for detection efficiency required for
performing a reconstruction, such as $50\%$ threshold for a
quantum homodyne tomography \cite{vogel,raymer}. Moreover, it
seems natural supposing that it is necessary to know precisely
what exactly one's measurement set-up detects (in more formal
language, it is seems necessary to know all the elements of the
positive valued operator measure, POVM, describing the
experiment).

Well, recently some works have appeared giving a rather obvious
hint: with quantum measurements it might be really unnecessary to
struggle for achieving exactly known (i.e. calibrated), and
perfectly controllable measurement set-up. Under certain quite
general conditions (such as, for example, Gaussianity of the state
is question) it is possible to update information about the set-up
and the signal state simultaneously (which was termed
"self-calibration") \cite{mogilevtsev2009}.  Very recently
experimental demonstrations of self-calibration were given
\cite{branczyk2012,kulik}. Moreover, it was demonstrated that
sufficient knowledge about set-up can be acquired in the process
of collecting data even in absence of any initial information
about the measurement set-up \cite{our fresh njp}.

However, self-calibrating approaches described above still suppose
that there is some fixed measurement set-up, albeit possibly
unknown one. But what if some noise is present? What if it is
subjected to random (and possible uncontrollable) change? In this
work we are demonstrating that classical noise affecting the
measurement set-up can be beneficial and usable for building
robust, efficient and simple measurement set-up (which can even be
much simpler than exiting methods) for certain tomography tasks.
Moreover, even randomly changing POVM parameters might be not an
obstacle for these tasks.

A general impression about a possible role of noise can be given
with the following simple example. Let us consider a set-up
performing projection on the coherent state with the amplitude
$\alpha$, i.e. with the POVM element
$\Pi=|\alpha\rangle\langle\alpha|$ (which can be realized with
heterodyne measurement \cite{yen}). If the amplitude $\alpha$
undergoes random changes (say, $\delta_j$) around some particular
value, say, $\alpha_0$, for a sufficiently large number of
different $\delta_j$ the resulting set of POVM elements (i.e. of
the form
$\Pi_j=|\alpha_0+\delta_j\rangle\langle\alpha_0+\delta_j|$) will
be sufficient for performing a complete state/process
reconstruction (which is attested by recent schemes of "data
pattern tomography" \cite{our fresh njp,our prl 2010} and quantum
process tomography \cite{lvovsky}). So, adding noise to a
measurement can actually increase a region that the measurement
set-up is actually "seeing", i.e. the search subspace.

Now let us demonstrate how noise can be implemented for devising
simple and efficient set-up for inference of photon number
distribution of a single-mode state of electromagnetic field.
Notice that, for example, for quantum homodyne tomography an
inference of photon number distribution is not much simpler than
inferring the complete density matrix (one needs performing the
same set of quadrature measurements). The task can be made easier
by building the specific set of POVM elements,
\begin{equation}
\Pi_j=\sum\limits_{n=0}^Nr_n|n\rangle\langle n|, \label{POVM1}
\end{equation}
where $|n\rangle$ are Fock states with $n$ photons. In the end of
90-s it was demonstrated that implementing only a single bucket
detector with a set of variable absorbers changing the efficiency
of the detection it is possible to collect data sufficient for
inferring the photon number distribution \cite{mog98}. For this
set-up one has $r_n=(1-\eta)^n$, where $\eta$ is the efficiency of
the detector. The method was realized experimentally \cite{paris}.
It was shown also that adding a coherent shift to the signal state
it is possible to perform a complete tomography \cite{us}.
However, an experimental realization of the scheme remained rather
challenging due to necessity to perform calibration of absorbers
for signals of a few-photon level. To avoid this more
sophisticated set-up was suggested and realized; there the signal
travels along the fiber loop splitting on each pass,
see\cite{loop,loopnew}).

\begin{figure}
\epsfig{ figure=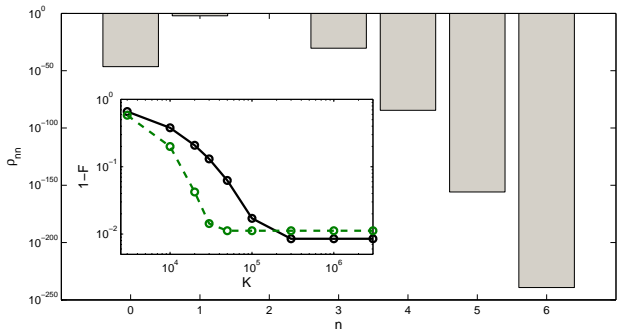,width=1.1\linewidth}
\caption { An example of two-photon Fock state maximal likelihood
estimation using variable thermal noise. 30 values of average
number,$\bar{n}$, of thermal photons were used equidistantly
distributed in the interval [0.1,0.95]; the search space was from
0 to 12 photons, 10,000 iterations of the reconstruction algorithm
were taken, starting with the maximal entropy state. The detector
efficiency, $\eta$, is 0.8. The inset shows a dependence of
fidelity on the number of iterations, $L$, for 20,000 measurements
(solid line) and 10,000 measurements (dashed line) per each value
of $\bar{n}$ \label{fig1} }
\end{figure}

Implementing noise gives a way to avoid using both sets of
calibrated absorbers or loop detection for the task. Indeed, let
us simply add a thermal noise (i.e. a source of detector "dark
counts") with the average number of photons $\bar n$ to our
signal. As follows from the Mandel's formula, without the thermal
noise mixed with the signal the probability to have no clicks on
our bucket detector is given by
\begin{equation}
p(\eta)=\sum\limits_{n=0}(1-\eta)^n\rho_{nn}, \label{prob1}
\end{equation}
where $\rho_{nn}$ are diagonal elements of the signal state
density matrix in the Fock state basis; notice that $p(\eta)$ is,
in fact, the photon-number generating function. When the thermal
light with the average number of photons, $\bar{n}$, is added to
the signal at the entrance of the detector, the probability
(\ref{prob1}) is modified as to \cite{perina,rockower}
\begin{equation}
p(\eta)\rightarrow
p(\eta,\bar{n})=\frac{1}{1+\eta\bar{n}}p\left(\frac{\eta}{1+\eta\bar{n}}\right),
\label{prob2}
\end{equation}
As it follows from Eq.(\ref{prob2}), one can just mix thermal
noises corresponding to different temperatures to build the POVM
elements set required for inference of photon-number distribution.
In Fig.\ref{fig1} an example of the two-photon Fock state
reconstruction is given. For the task 30 different values of the
average number of photons of the added thermal noise are taken.
Efficient expectation-maximization iterative algorithm for maximal
likelihood estimation was implemented \cite{us,paris}. For 10,000
measurements per each value of $\bar{n}$ a fidelity of more than
95$\%$ was reached for 10,000 iterations starting from the maximal
entropy state. One should point out two important features of the
suggested scheme. The first one is necessity to use rather large
search subspace due to the photon-number distribution of the
thermal state being long-tailed. The second one is the slower
convergence of the algorithm for larger number of measurements
(see inset in Fig.\ref{fig1}). Increasing of the number of
measurement ultimately is eventually leading to increase in the
reconstruction accuracy. However, with larger number of
measurements one can actually get worse results for the same
number of iterations.

It should be stress out that adding thermal noise leads to
significant simplification of the process of building the POVM
elements set required for the reconstruction. Indeed, one is not
even obliged to actually change parameters of the source of
thermal noise; it is sufficient to change a time-window of
detection to change effectively an average number of thermal
photons. Notice, that thermal noise needs not to be
pre-calibrated: it is possible to change noise temperature
arbitrarily and calibrate it by temporarily switching off the
signal and collecting data. It means that one can actually perform
the reconstruction with usual daylight. The only obstacle is
dispersion of the detection efficiency, so it is necessary to
filter thermal noise to provide for the constant detector
efficiency in the spectral interval of thermal noise and the
signal.

Of course, when one tries to build set of POVM by adding noise,
the question arises of its influence on possible reconstruction
errors. Generally, the problem of error estimation for quantum
tomography is rather "hot" and controversial subject nowadays.
Both method bases on ML estimation and Bayesian inference were
suggested for the purpose (see, for example, Ref.\cite{cristandl}
and references therein). However, for diagonal elements inference
with general POVM (\ref{POVM1}) we can suggest a simple estimation
of error of the ML method along the lines suggested in
Ref.\cite{mognjp} (and somewhat in the spirit of approach used in
Ref.\cite{cristandl}).

\begin{figure}
\epsfig{ figure=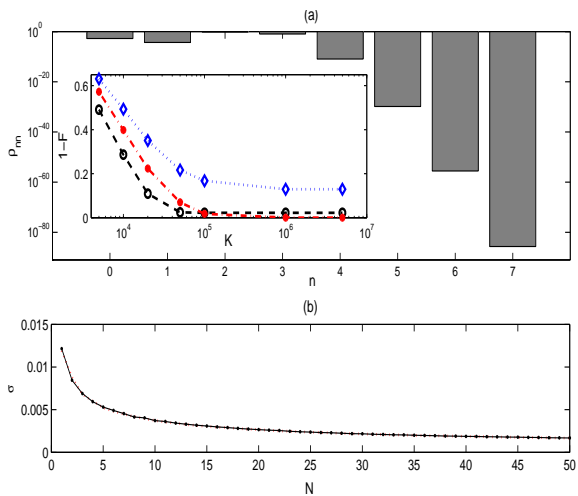,width=1.1\linewidth,height=6.5cm}
\caption { An example of two-photon Fock state maximal likelihood
estimation using random variable thermal noise (a). The noise was
taken to be Gaussian;  30 values of average number,$\bar{n}$, of
thermal photons were used equidistantly distributed in the
interval [0.1,0.95]; the variance is 0.1; the search space was
from 0 to 12 photons, 1,000,000 iterations of the reconstruction
algorithm were taken, starting with the maximal entropy state. The
detector efficiency, $\eta$, is 0.8; the number of measurements is
50,000 per each value of $\bar{n}$. The inset shows a dependence
of fidelity on the number of iterations, $K$, for 50,000
measurements and variance 0.1 (dash-dotted line); 10,000
measurements and variance 0.01 (dashed line); 10,000 measurements
and variance 0.1 (dotted line) per each value of $\bar{n}$. The
panel (b) shows an estimation of maximal errors made according to
Eq.(\ref{error}) for different number of trials, $N\times1000$ per
average value (solid line with dot marks), in comparison with the
estimation for non-random POVM for fixed noise values equal to
average ones. For this figure  the average number of thermal
photons is $\bar{n}=0.1$; the variance is 0.1; the search space
was from 0 to 12 photons (dotted line). \label{fig2} }
\end{figure}

Let us consider our measurement with the general POVM
(\ref{POVM1}) as the set of measurements with $K$ complete POVMs
each having just two elements $\Pi_j$ and $I-\Pi_j$, $j= 1,\ldots,
K$. In the limit of large number $N_j\gg1$ of runs with the $j$-th
POVM of the set, the likelihood function has the following form
\begin{eqnarray}
& & P(S_1,S_2,\ldots,S_K|\rho) \nonumber\\
& & =
\prod_{j=1}^K\frac{N_j!}{S_j!(N_j-S_j)!}\mathrm{tr}(\Pi_j\rho)^{S_j}\mathrm{tr}([I-\Pi_j]\rho)^{N_j-S_j},\qquad
\label{E1}\end{eqnarray} with the number of outcomes $S_j $
corresponding to the POVM element $\Pi_j$ (and, respectively,
$N_j-S_j$ for $I-\Pi_j$), can be approximated by using the
well-known large-$N$ limit of the binomial distribution (see, for
instance, Ref. \cite{Gnedenko}). The latter is a Gaussian
approximation, which in our case gives
\begin{eqnarray}
& & P(S_1,S_2,\ldots,S_K|\rho)
\nonumber\\
& & \approx \prod_{j=1}^K \frac{1}{\sqrt{2\pi
N_j}\sigma_j}\exp\left\{-N_j \frac{
\left[\frac{S_j}{N_j}-\mathrm{tr}(\rho\Pi_j)\right]^2}{2\sigma_j^2}\right\},
\label{E2}
\end{eqnarray}
with\[\sigma^2_j =
\mathrm{tr}([I-\Pi_j]\rho)\mathrm{tr}(\Pi_j\rho).\] For
sufficiently large $N_j$ the width of the peak around the maximum
likelihood point, i.e. the admissible values for $\rho_{mm}$,
$m=0,1,\ldots, M$,  is given by the variance appearing in Eq.
(\ref{E2}).  Therefore, the   error $\delta$ for each $\rho_{mm}$
of the maximum likelihood estimate is  on the order
\begin{equation}\delta \sim
\mathrm{max}\frac{\sigma_j}{\sqrt{N_j}}=\mathrm{max}\left\{\sqrt{\frac{S_j}{N^2_j}(1
- \frac{S_j}{N_j})}\right\}.\label{error}\end{equation} An example
of maximal error estimation for the two-photon Fock state
reconstruction is shown in Fig. \ref{fig2}(b). One can see that
Eq.(\ref{error}) gives rather reasonable estimate of a maximal
error for different number of measurements.

The most obvious prerequisite for the "thermal noise"
reconstruction discussed above seems to be a constant predefined
level of the noise. However, it is not hard to see that noise can
be varied during the data collecting process. It can be even
random. Indeed, if the set of parameters, $\vec{\mu}$, describing
a particular POVM element in Eq.(\ref{POVM1}), $\Pi(\vec{\mu})$,
represents a continuous random variable, then for the probability
one has simply
\begin{equation}
p=\sum\limits_{n=0}[\bar{r}]_{n}\rho_{nn}, \quad
[\bar{r}]_{n}=\int d\vec{\mu} [r(\vec{\mu})]_{n}, \label{average
povm}
\end{equation}
Of course, for each run of the experiment the result will depend
on the particular value of random parameters, $\vec{\mu}$. The
reconstruction procedure hinges on the fact that for the
sufficiently long sequence of trials one can assume the all
results were obtained with the averaged POVM elements
(\ref{average povm}).  For our example of the "thermal noise"
reconstruction these average POVM elements are to be defined on
the calibration stage preceding a measurement of the state mixed
with noise. Fig.\ref{fig2}(a) demonstrates that the reconstruction
with "noisy" POVM elements is indeed feasible. There an example of
two-photon state inference is shown for the "thermal noise"
reconstruction scheme with average number of photons fluctuating
with the normal distribution. It is remarkable that one is able to
achieve quite accurate reconstruction results with rather strong
noise (when the variance of noise is comparable with the average
value of $\bar{n}$). The pay-off is the necessity to increase the
number of measurements per assumed averaged POVM element. However,
this increase is not crucial (for instance, just five times for
the example shown in Fig.\ref{fig2}, corresponding to quite large
variance of the noise). An estimation (\ref{error}) shows that
even for  moderate number of trials maximal errors for fixed
thermal noise and random one can be quite close (this situation is
illustrated in the Fig.\ref{fig2}(b)). Also, as the inset in
Fig.\ref{fig2} demonstrates, random variation of the thermal noise
do not noticeably worsen convergence of the reconstruction
procedure in comparison with the "fixed noise" results shown in
Fig.\ref{fig1}.

\begin{figure}
\epsfig{ figure=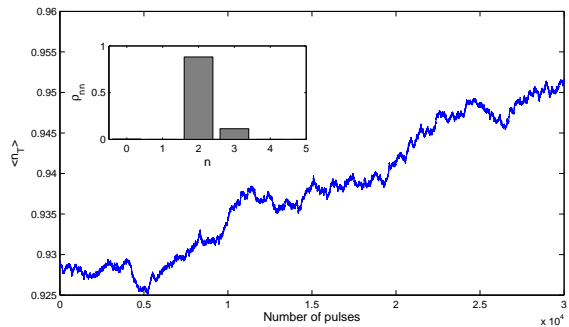,width=\linewidth}
\caption { An example of a realization of the randomly changing
average number of thermal photons. Random change is 1D random walk
with probability 0.51 of jumping to a higher value of $\bar{n}$
(they are taken to be a discrete set with the step $5*10^{-4}$.
The total of $9*10^8$ measurements were assumed. For the purpose
of the reconstruction this set was divided on 30 subsets with
30,000 measurements in each. So, 30 different POVM elements
(\ref{prob1}) were assumed for the reconstruction with different
$\bar{n}$ corresponding to averages over each subset. $10^6$ steps
of the iteration procedure were takes. The inset shown the
convergent result of the estimation of the signal two photon
state.) \label{fig4} }
\end{figure}

So, we have established that random variations around some fixed
values of POVM parameters do not spoil the reconstruction. Now let
us demonstrate that we can relax our requirement for controlling
the experiment even further. Values of POVM parameters might be
not fixed at all. A knowledge of average values of these
parameters in a certain time interval is sufficient for performing
the reconstruction. In Fig.\ref{fig4} an example of the
reconstruction is shown for the average number of photons of the
mixed thermal state undergoing the stochastic drift process. Since
copies of the signal are assumed to be generated as an equidistant
sequence of pulses, random change of the average number of photons
was taken to be described as 1D random walk process. Probability
of jump to higher $\bar{n}$ was taken to be slightly larger, than
to lower $\bar{n}$ ($1\%$ higher), so the average number of
photons was gradually increasing (see Fig.\ref{fig4}). Consequent
sets of 30,000 measurements were taken for averaging and assigning
an averaged POVM element (whole of 30 different values of
$\bar{n}$ non-equdistantly distributed in the interval
[0.1,0.952]). As it can be seen in Fig.\ref{fig4}, this
approximation has allowed to achieve rather good estimation of the
two-photon signal state even with rather moderate number of
measurements.

Concluding: we have established that noising the measurement (or
adding noise to the signal) can be a simple and efficient way to
produce a set of POVM elements sufficient for reconstructing the
state of the signal. Mixing the signal with the thermal noise on
the detector (i.e. adding "dark counts"), one can make very simple
and non-expensive set-up for estimating a photon number
distribution. One does not require for it a set of calibrated
absorbers or loop detectors. Calibration of the thermal noise can
be done directly in the process of measurement. Randomness of the
added noise does not spoil the reconstruction, provided that the
average values of noised parameters are known for sufficiently
long series of trials. We have demonstrated that both for random
parameters fluctuations around some set of fixed values,  and for
the random change similar to stochastic drift/random walk. The
latter feature hints at the possibility to overcome the main
reason of using the same source for generating the supposedly
unknown signal and the reference field in the schemes of complete
quantum tomography: the phase drift. Thus, this work is the step
to reaching the ultimate goal of a quantum state tomography:
accomplishing the reconstruction of truly unknown signal state.

The authors are thankful for Prof. Jan Perina for fruitful
discussions. This work was supported by Foundation of Basic
Research of the Republic of Belarus, by the National Academy of
Sciences of Belarus through the Program "Convergence", by the
Brazilian Agency FAPESP (project 2011/19696-0) (D.M.);  and has
received funding from the European Community's Seventh Framework
Programme (FP7/2007-2013) under grant agreement n$^\circ$ 270843
(iQIT) (N.K.).


\begin{thebibliography}{99}

\bibitem{all} M. G. A. Paris and J. \v{R}eh\'{a}\v{c}ek (Eds), \emph{Quantum
states estimation}, Lect. Notes Phys. vol. 649 (Springer, Berlin
Heidelberg, 2004).

\bibitem{Ourjoumtsev} A. Ourjoumtsev, R. Tualle-Brouri, J. Laurat, P. Grangier, Science
312, 83 (2006).

\bibitem{vogel} K. Vogel and H. Risken, Phys. Rev. A\textbf{40},
2847 (1989).

\bibitem{raymer} D. T. Smithey, M. Beck, M. G. Raymer, and A. Faridani, Phys.
Rev. Lett. \textbf{70}, 1244 (1993).

\bibitem{mogilevtsev2009} D. Mogilevtsev, J. Rehacek, and Z.
Hradil, Phys. Rev. A \textbf{79},  (2010) 02010(R); D.
Mogilevtsev, Phys. Rev. A \textbf{82},  (2010) 021807(R); D.
Mogilevtsev, J. Rehacek, and Z. Hradil, New J.Phys.  \textbf{14},
(2012) 095001.

\bibitem{branczyk2012} A. M. Branczyk, D. H. Mahler, L. A. Rozema, A. Darabi, A. M. Steinberg and
D. F. V. James, New J.Phys.  \textbf{14},  085003 (2012).

\bibitem{kulik} S. Straupe, D. Ivanov, A. Kalinkin, I. Bobrov, S. P. Kulik, D. Mogilevtsev,
arXiv:1112.3806v2 (2013).

\bibitem{our fresh njp} D. Mogilevtsev, A. Ignatenko, A. Maloshtan,
B. Stoklasa, J. Rehacek and Z. Hradil, Data pattern tomography:
reconstruction with unknown apparatus, to appear in New J. Phys.
(2013).

\bibitem{yen} H. P. Yuen and J. H. Shapiro, IEEE Trans. Inf. Theory \textbf{24}, 657
(1978); \textbf{25}, 179 (1979).

\bibitem{our prl 2010} J. Rehacek, D. Mogilevtsev, and Z. Hradil, Phys. Rev. Lett.
\textbf{105}, 010402 (2010).

\bibitem{lvovsky} M. Lobino, D. Korystov, C. Kupchack, E. Figueroa, B. C. Sanders and A. I. Lvovsky, Science \textbf{322}
563(2008); S. Rahimi-Keshari, A. Scherer, A. Mann, A. T.
Rezakhani, A. I. Lvovsky and B. C. Sanders, New J. Phys.
\textbf{13} 013006 (2011).

\bibitem{mog98} D. Mogilevtsev, Opt. Comm. \textbf{156}, 307 (1998);
D. Mogilevtsev, Acta Physica Slovaca \textbf{49}, 743 (1999).

\bibitem{paris} A.R. Rossi, S. Olivares, M.G.A. Paris, Phys.
Rev. A \textbf{70}, 055801 (2004); A.R. Rossi and M.G.A. Paris,
Eur. Phys. J. D \textbf{32}, 223 (2005); G. Zambra, A. Andreoni,
M. Bondani, M. Gramegna, M. Genovese, G. Brida, A. Rossi, and
M.G.A. Paris, Phys. Rev. Lett. {\bf 95} 063602 (2005).

\bibitem{us}  Z. Hradil, D. Mogilevtsev, and J. \v{R}eh\'{a}\v{c}ek,
Phys. Rev. Lett. \textbf{96}, 230401 (2006); D. Mogilevtsev, J.
\v{R}eh\'{a}\v{c}ek and Z. Hradil, Phys. Rev. A \textbf{75},
012112 (2007).

\bibitem{loop} J. Rehacek, Z. Hradil, O. Haderka, J. Perina Jr, M. Hamar,
Phys. Rev. A \textbf{67}, 061801(R) (2003); O. Haderka, M. Hamar,
J. Perina Jr, Eur. Phys. J. D \textbf{28}, 149 (2004).

\bibitem{loopnew} J. G. Webb and E. H. Huntington, Opt. Exp. \textbf{17} 11799 (2009).

\bibitem{perina} J. Perina, \emph{Quantum Statistics of Linear and Nonlinear Optical
Phenomena}, (Springer; Berlin, Heidelberg; 2nd rev. ed., 1991).

\bibitem{rockower} E. B. Rockower, Phys. Rev. A \textbf{37}, 4309
(1987).


\bibitem{mognjp} Z. Hradil, D. Mogilevtsev, and J. Rehacek,
New J. Phys. \textbf{10}, 043022 (2008).

\bibitem{cristandl} M. Christandl and R. Renner, Phys. Rev. Lett. \textbf{109}, 120403
(2012).

\bibitem{Gnedenko} B. V. Gnedenko,  \textit{The Theory of Probability} (English Translation; Mir Publishers, Moscow, 1978), p. 85.

\end{thebibliography}
\end{document}